\documentclass[12pt]{iopart}
\usepackage{graphicx,psfig,epsfig}
\def\beq{\begin{equation}}
\def\eeq{\end{equation}}
\def\ba{\begin{eqnarray}}
\def\ea{\end{eqnarray}}
\begin{document}

\title[Properties of quark matter produced in heavy ion collision]
{Properties of quark matter produced in heavy ion collision}
\author{J. Zim\'anyi\dag, P. L\'evai\dag, T.S. Bir\'o\dag\ddag}
\address{\dag\  RMKI Research Institute for Particle and Nuclear Physics, \\
 P.O. Box 49, Budapest H-1525, Hungary}
\address{\ddag\ Institute for Theoretical Physics, University of Giessen, \\
 16 Heinrich-Buff-Ring, Giessen, D-35392, Germany}

\ead{jzimanyi@rmki.kfki.hu}

\begin{abstract}
We describe the hadronization of quark matter
assuming that quarks creating hadrons coalesce from a
continuous mass distribution.
The pion and antiproton spectrum as well as the momentum dependence of the 
antiproton to pion ratio are calculated.
This model reproduces fairly well the experimental data at RHIC energies.
\end{abstract}
\submitto{\JPG}

\section{Introduction}

The most important aim of heavy ion experiments performed at the 
RHIC accelerator besides the production of quark matter is the 
reconstruction of its properties from the observed hadron measurables. 

The investigations of this problem are focused around three main questions: 
a) the effective mass of the quarks, 
b) the momentum distribution of quarks, 
c) the hadronization mechanism.

Regarding the quark mass in quark matter 
up to now essentially two different assumptions has been made:
i) the quark matter consists of an ideal gas of
 non-interacting  massless quarks and gluons and
this type of quark matter undergoes a first order phase transition
during the hadronization; or 
ii) the effective degrees of freedom in quark matter
are dressed quarks with an effective mass of about $ 0.3$ GeV, which  mass
is the result of the interactions within the quark matter; this type of
quark matter clusterizes and forms hadrons smoothly, following a cross-over 
type deconfinement -- confinement transition. 

Considering the momentum distribution in quark matter one meets the 
following ideas:
The bulk part of the quarks is produced in soft processes,
these quarks have a fairly exponential transverse momentum distribution
on the top of a collective transverse flow in heavy ion collisions. 
On the other hand, quarks with large transverse momentum 
($p_T \geq 2-4$ GeV) are viewed as produced mainly in hard collisions and
their spectra obey a power law shape described by the functional form
of fragmentation functions.

Finally one meets the worthy problem of hadronization. 
In early publications~\cite{Muller80,BiZi}
the hadronization was assumed to happen through a quasi-static,
first order phase transition. However, as the experiments 
have suggested that the whole lifetime of the heavy ion collision
is very short, other non-equilibrium dynamical models
obtained more and more credit. One of the most successful description
is based on quark coalescence 
\cite{ALCOR,ALCOR00,recHwa,coalGKL,recFries,coalMolnar}. 
This idea is supported by recent RHIC experiments~\cite{STARcoal}.

In this paper we present a new model of the late phase
of the heavy ion collision. 
In this approach the effective quarks are massive and
have a particular mass distribution described in Section 2. 
In Section 3 the quark momentum distribution is studied in
details. Section 4 summarizes the appropriate expressions for
the coalescence processes in this environment.
Numerical results and comparison to experimental data are
displayed in Section 5. We discuss the obtained results in Section 6.

\section{Mass distribution of Quarks}

We assume that in hot and dense deconfined
matter the average interaction of the massless (current) quarks generate
an effective mass for them. Several analyses of lattice QCD calculations
supports the appearance of massive quark degrees of
 freedom~\cite{LevHeinz,Kampf}.
(According to these analyses the gluons become very heavy just before 
hadronization, and thus we neglect their explicit degrees of freedom
in the models.)
The mass of these quasi-particles is in the range of $0.3-0.4$ GeV around
the phase transition temperature.
In the early approaches the quasi-quarks have a sharp, momentum 
independent mass.  
In a realistic approach, however, there is no reason to assume that 
the interacting quarks behave as on-shell particles. Due to the residual
interaction the  mass of  the  quasi-quarks may be distributed in an 
interval of about a few hundred MeV while having 
a maximum in the range of $0.3-0.4$ GeV. 
The shape of this distribution is presently
not known 
\footnote{Not thinking that quarks are dressed only by one
 partner particle
(gluon), there is also no reason to stick with the
 Breit-Wigner form for this distribution.}. 
One may want to investigate the effect of different spectral functions 
on the hadronization process.  In the present work
we consider the following mass distribution:

\begin{equation}
   \rho(m) = {\it N} e^{ - \frac{\mu}{T_c} \sqrt{\frac{\mu}{m}+\frac{m}{\mu}} }
\label{rhom}
\end{equation}

 \begin{figure}[htb]
 \begin{center}
\resizebox{100mm}{100mm}{\includegraphics{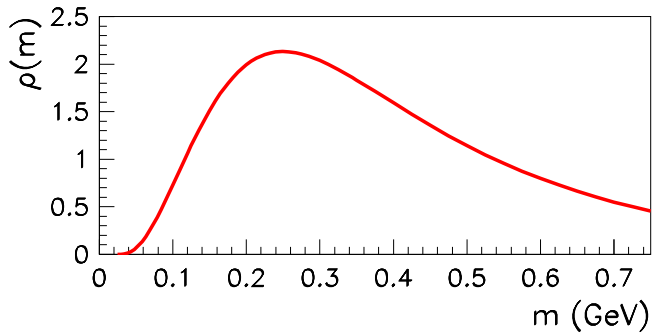}}  
\vspace*{-4.3truecm} 
 \caption{Distribution of the effective quark mass in eq.~(\ref{rhom}).}
 \label{figure1}
 \end{center}
 \end{figure}

This distribution vanishes both for too small and too large masses and has a
smooth maximum around the familiar valence quark mass value ($ \mu $).
Fig.\ref{figure1} displays this mass distribution for $\mu = 0.25 $ GeV
and $T_c=0.26$ GeV.

\section{ Momentum distribution of quarks  } 

In the RHIC experiments it was observed, that the  
produced particles show a nearly exponential 
spectrum at low transverse momenta and a power-law like one at
high transverse momenta. This distribution
was considered in earlier publications as a sum of three
components produced by different physical mechanisms: 
a) thermal equilibrium, b) coalescence,
c) fragmentation. The relative weights of these components were determined
by a fit to the experimental data.
In the present work we assume  an analytic, parameterized 
transverse momentum distribution,
which is similar to an exponential distribution for low momenta
and a power law distribution for large momenta.
The phase space density of effective quarks with a variable mass $m$,
transverse momentum $p_T$, coming from an angular direction $\phi$ and
a coordinate rapidity $\eta$ is given by 

\beq
f_{quark}(p_{\perp},m,\phi,\eta) =
\frac{ A }{ [1+ (q-1)*\frac{E(p_{\perp},m,\phi,\eta)}{T} ]^{(q/(q-1))} },
\label{fp}
\eeq 

Here $q$ is a  parameter 
characterizing a possible deviation from the usual Boltzmann distribution,
which is recovered for $q=1$.
The parameter $T$ is in general not  the usual temperature, but in
the Boltzmann limit, $q \rightarrow 1 $ it becomes the canonical temperature.
That which values of $q$ and $T$ fit experimental heavy ion data best,
may depend on the transverse momentum range considered: a $q$ value not
equal one fits usually to much higher momenta than the original Boltzmannian
assumption.

Although the expression in eq.~(\ref{fp})  coincides with the form
of the Tsallis distribution~\cite{Tsallis} (as many simple minded fits do),  
we do not necessarily have to assume  
a special equilibrium state, which is considered in the  
non-extensive thermodynamics.  
One may assume such an equilibrium Tsallis system in quark matter
if one of the following conditions is fulfilled:
i) there is an energy dependent random noise in the system not 
reflecting a constant temperature;
ii) the system contains a finite number of particles which interact 
indefinitely long; 
iii) there are long range forces in the system. 
We realize, that such 
conditions may be found in a fireball produced in high energy heavy ion
collisions, it is just not sure, whether the time is enough to arrive at 
such a special equilibrium state.

The energy $ E $, appearing in  
eq.(\ref{fp})  is given by the blast wave formula: 
\beq 
E(p_{\perp},m,\phi,\eta) = \gamma
 ( \sqrt{m^2 + p_{\perp}^2 }\cdot \cosh(\eta) - v_{flow} \cdot p_{\perp}
 \cdot \cos(\phi) ).
\label{en}
\eeq
assuming a homogeneous transverse flow velocity, $v_{flow}$, with a 
corresponding Lorentz factor, $\gamma$.
In eq.(\ref{en}) the effect of an average "blue shift", caused
by the hydrodynamical flow, is taken into account.

\section{Coalescence of Massive Quarks}

Using the covariant coalescence model of Dover {\it et al.}~\cite{dover},
the spectra of hadrons formed from the coalescence of quark and antiquarks
can be written as

\begin{equation}
 E \cdot \frac{dN_h}{d^3p} = \frac{dN_h}{dy p_T dp_T d\phi_p} =
 \frac{g_h}{(2\pi)^3}  \int (p_h^{\mu} \cdot d\sigma_{h,\mu}) 
  F_h(x_h;p_h).
\label{coal1}
\end{equation} 

Here $F_h(x_h;p_h)$ is an in principle eight dimensional distribution
(Wigner function) of the formed hadron.
Furthermore $d\sigma$ denotes an infinitesimal element of the space-like 
hypersurface where the hadrons are formed, while
$g_h$ is the combinatorial factor for forming a colorless hadron
from a spin 1/2 color triplet quark and antiquark.
For pions the statistical is $ g_{\pi} = 1/36 $ , for antiprotons 
$ g_{\overline p} = 1/108 $ \cite{coalGKL}.

Assuming  an instant hadronization at a sharp proper time,
$ \tau $,  with cylindrical symmetry  one gets
\begin{eqnarray}
 \frac{dN_h}{dy p_T dp_T d\phi_p} =
   \frac{g_h}{(2\pi)^3}  &\int& \tau \cdot m_{\perp} cosh(y-\eta) 
   d\eta r dr d\phi F_h(x_h;p_h) \nonumber 
\label{coal12}
\end{eqnarray} 
\vspace{0.5cm}  
Here $y$ is the momentum rapidity of the hadronic four-momentum 
$p_h$ and $\eta$ is the coordinate-rapidity of the formation point. 
For $y=0$ and $\phi=0$  eq.(\ref{coal1})becomes
\begin{equation}
\left. \frac{dN_h}{dy p_T dp_T d\phi_p}\right|_{y=0} = 
\frac{g_h}{(2\pi)^3}\cdot \pi R^2 \tau \cdot m_{\perp} \cdot
\int_{\Delta_{\eta}/2}^{\Delta_{\eta}/2}\!\!d\eta \: \cosh(\eta) 
\: F_h(p_h;\eta) 
\label{coal2}
\end{equation}

Assuming that we get contribution to the spectra mainly from the
$ \Delta_{\eta} << 1 $ region, eq.(\ref{coal2}) simplifies to
\begin{equation}
\left. \frac{dN_h}{dy p_T dp_T d\phi_p}\right|_{y=0} \:  = \:
\frac{g_h}{(2\pi)^3} \cdot \pi R^2 \tau \Delta_{\eta} \cdot m_{\perp} \cdot
 F_h(p_h;0)
\label{coal3}
\end{equation}
where $m_{\perp}=\sqrt{m^2+p_T^2}$ is the transverse mass, 
$\phi_p$ the angle of the
transverse momentum of the hadron and $R$ is the cylindrical radius.
Introducing the  hadronization volume, $  V = \Delta_{\eta} \tau R^2 \pi$, 
and the abbreviation $F_h(p_{\perp},0) = F_h(p_{\perp})$ we finally get
\begin{equation}
\left. \frac{dN_h}{dy p_T dp_T d\phi_p}\right|_{y=0} \:  = \:
 \frac{g_h}{(2\pi)^3} V  m_{\perp} F_h(p_{\perp}) 
\end{equation}

Assuming that the mesons are produced by coalescence of
particle {\it a } with mass $m_a$ and particle {\it b } with
mass $m_b$ , the meson source function, $ F_M({\bf p}_M) $,
will have the form:
 
\begin{equation}
   F_M({\bf p}_h) = \int 
  d^3{\bf p}_a  d^3{\bf p}_b     
 f_a({\bf p}_a;0)f_b({\bf p}_b;0)C_M(p_a,p_b,p_M).
\label{coal4}
\end{equation}

In eq.(\ref{coal4}) 
the coalescence function $C_M(p_a,p_b,p_M)$  is the probability
for particles  with four momenta $ p_a, p_b$ to form a
meson with four momentum $p_M $. We assume a Gaussian wave-packet shape
being maximal at zero relative momentum (characteristic for s-wave internal 
hadronic wave functions): 
\begin{equation}
 C_M(p_a,p_b,p_M) = \alpha_M \cdot  e^{-((p_a - p_b)/P_c)^2}  
\label{coalfv}
\end{equation}

The parameters $\alpha_M$ and $P_c$ reflect properties of the hadronic
wave function in the momentum representation convoluted with the formation
matrix element.
For the present consideration we assume that the system consist of a mixture
of effective particles having different masses. The densities of
 particles with mass m are weighted with the corresponding mass
 distribution function, $\rho(m)$, eq.(\ref{rhom}). 

First we deal with the formation of mesons.
We assume that the total mass and momentum of a hadron will be combined
additively from that of the constituents, but $P_c$ is so small, that
that practically zero relative momentum partons form a meson. 
This leaves us with
\begin{eqnarray}
   {\bf p_a} = {\bf p_b} &=& {\bf p_M} / 2  \\ 
             m_a + m_b &=& m_M  
\label{cons2}
\end{eqnarray}
This  leads to the coalescence function
\begin{equation}
   C_M = \alpha_M \cdot \delta(\vec{p}_a-\vec{p}_M/2)
 \cdot \delta(\vec{p}_b-\vec{p}_M/2) \cdot \delta(m_a+m_b - m_M)  
\end{equation}.
Thus we arrive at the following meson distribution function:  
\begin{eqnarray}
 F_M({\bf p_t};0) = \alpha_M &\cdot& \int_0^{m_M} dm_a
 \int_0^{m_M} dm_b \cdot \delta(m_M -(m_a+m_b)) \nonumber \\
 &\cdot& \rho(m_a) \cdot f_q({\bf p_t}/2;m_a)  \nonumber \\
 &\cdot& \rho(m_b) \cdot f_b({\bf p_t}/2;m_b).
\label{coalM}
\end{eqnarray} 
\vspace*{-1.8truecm} 
 \begin{figure}[htb]
 \begin{center}
\resizebox{100mm}{100mm}{\includegraphics{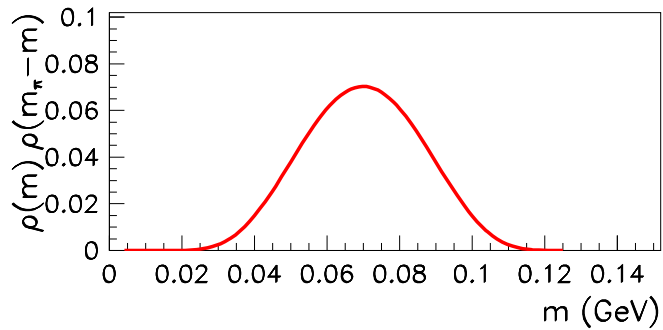}}
\vspace*{-4.8truecm} 
 \caption{The product of the two quark-mass distributions,
          $\rho(m) \cdot \rho(m_{\pi} - m)$, in eq(\ref{coalM}).
  The maximum of the product is at $m=m_{\pi}/2$.
}
 \label{figure2}
 \end{center}
 \end{figure}

Fig.\ref{figure2}. shows the convolution of two quark-mass distribution
functions appearing in the coalescence probability eq(\ref{coalM}) for
the coalescence production of a pion.

One observes, that the product of the two
mass distributions always has a  maximum around $m= m_{\pi}/2$. 
Already the naive coalescence approach works, whenever this maximum is
sharp enough.  

Similar argumentation leads to the three-fold coalescence expression. 
In the place of
 eq.(\ref{cons2}) we get:
\begin{eqnarray}
   {\bf p_1} ={\bf p_2}={\bf p_3} &=& {\bf p_B} / 3  \\
     m_1 +m_2 +m_3 &=& m_B
\end{eqnarray} 
and the baryon source function becomes
\ba
  F_{B}(p_t)
     &=& \alpha_B \cdot \int_0^{m_{pr}}\!\!\!dm_1 \int_0^{m_{pr}}\!\!\!dm_2 \int_0^{m_{pr}}\!\!\!dm_3 \nonumber \\
     &.&   \rho(m_1) \, f_{q}(p_t/3,m_1)) \nonumber \\
     &.&   \rho(m_2) \, f_{q}(p_t/3,m_2))   \nonumber \\
     &.&   \rho(m_3) \, f_{q}(p_t/3,m_3))   \, \delta( m_B - (m_1+m_2+m_3) ) .     
\label{coalB}
\ea

 \begin{figure}[hb]
 \begin{center}
\rotatebox{270}{\includegraphics[width=100mm,height=150mm]{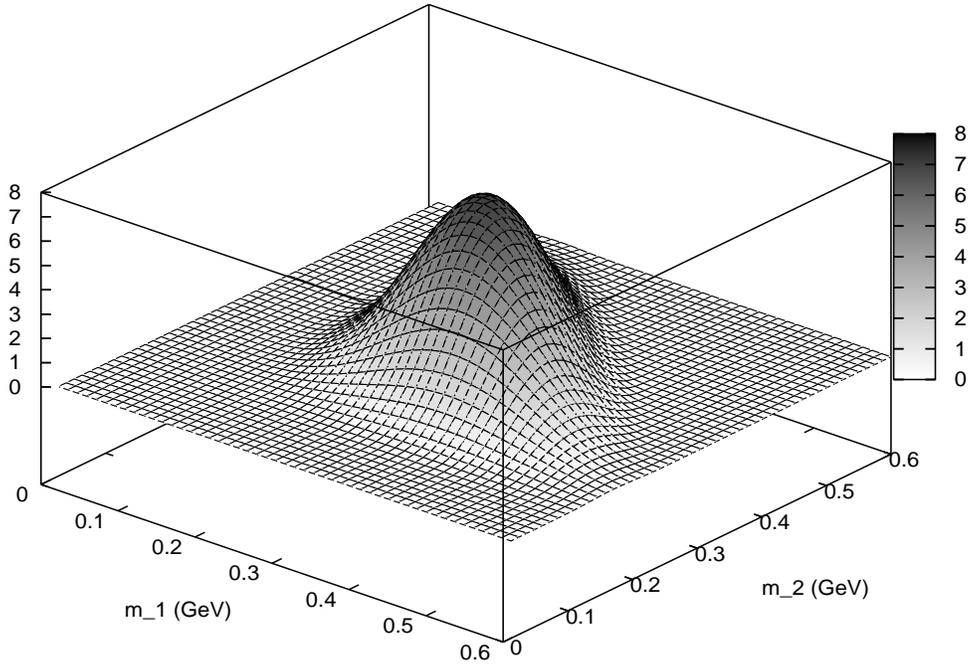}}
 \caption{Product of three quark-mass distributions,
          $\rho(m_1)\cdot \rho(m_2) \cdot \rho(m_3)$, from eq.~(\ref{coalB}).
          Horizontal axes connect to masses  $m_1$ and $m_2$.
          Mass  $m_3$ is defined as $m_3 = m_{\overline p}-(m_1+m_2)$.
}
 \label{3dm}
 \end{center}
 \end{figure}

Fig.\ref{3dm}. shows the product of the three quark-mass distribution 
functions appearing in eq.(\ref{coalB}).

\newpage
Figures 2 and 3 shows that the maximum contribution to the 
coalescence integrals are obtained from the equal
mass part of the mass distributions.

At the end of this section we may compare the basic philosophy of
the different coalescence models. In all cases we met 
the problem of mismatch of initial and final state
quantum numbers. To resolve this problem different solutions
are proposed:

a) In the quantum mechanical approach the overlap of initial 
and final state wave function is calculated. In this process - 
simply said - the quantum mechanical uncertainty bridges
the mismatch gap.

b) In the most often used classical approximation the coalescence 
probability is smeared around the momentum
 conservation, Ref.~\cite{dover}.

c) In the present approximation in the coalescence transition the
momentum is strictly conserved, as in Ref.~\cite{recFries}, but the
masses of the particles of the initial state have a finite
width distribution.

\section{Numerical results and comparison with experimental data } 

\begin{figure}[htb]
 \begin{center}
\resizebox{100mm}{90mm}{\includegraphics{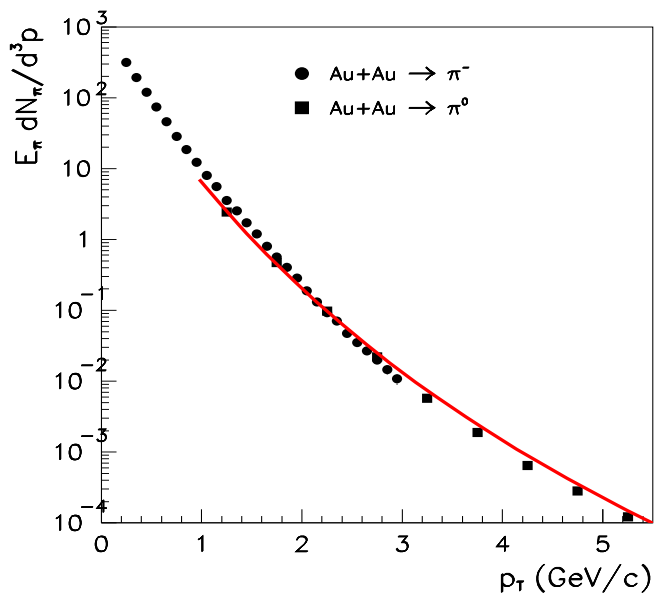}}
 \caption{Measured pion spectra 
 in central Au+Au collisions at $\sqrt{s}=200$ AGeV~\cite{PHENIX}.
 Full line indicates the calculated pion yield.
}
 \label{figpi}
 \end{center}
 \end{figure}

The parameters used in the calculations are as follows. 
\noindent
$ \mu =0.25$ GeV, $T_c=0.26$ GeV, 
$ T =0.07$ GeV, $ q =1.19$, $ v_{flow} =0.56 $. 
The $ \alpha_M $ and $ \alpha_B $ coalescence constants were adjusted to 
fit the absolute value of the measured $\pi$ and $ \overline{p} $ yields.
In Fig.(\ref{figpi}) both the $ \pi^- $ and $\pi^0 $ data are displayed.

In  the Figures 4-6 we show the calculated distributions together with
the corresponding experimental data. The experimental points are
taken from Ref.~\cite{PHENIX}.

\vspace*{-0.1truecm}

\vspace*{-1.0truecm}
 \begin{figure}[htb]
 \begin{center}
\resizebox{100mm}{90mm}{\includegraphics{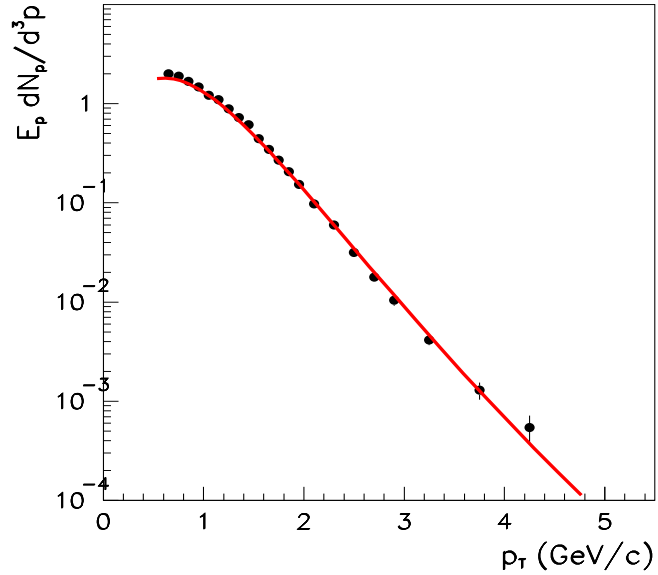}}
 \caption{Transverse momentum spectrum for antiproton production
 in central Au+Au collisions at $\sqrt{s}=200$ AGeV~\cite{PHENIX}.
Full line indicates our calculation.
}
 \label{figpr}
 \end{center}
 \end{figure}

\vspace*{-1.0truecm}
\begin{figure}[htb]
 \begin{center}
\resizebox{100mm}{90mm}{\includegraphics{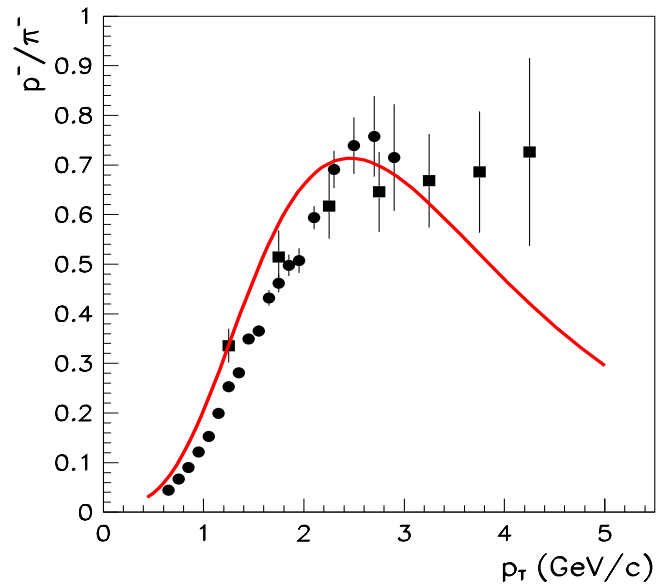}}
 \caption{The ${\overline p}/\pi^{-}$ ratio as a function of $ p_T$
on a linear scale in central Au+Au collisions 
at $\sqrt{s}=200$ AGeV~\cite{PHENIX}.
}
 \label{figrat}
 \end{center}
 \end{figure}

\vspace*{1.0truecm}
\section{Conclusion}

In earlier publications~\cite{coalGKL,recFries} 
it was assumed that for the low, medium and high 
transverse momentum intervals the hadrons are produced with different
mechanisms, via thermal, coalescence and fragmentation processes.
In the present work we proposed a model, in which, besides the
momentum, the energy is also conserved to a good approximation 
in the coalescence process. 
This model yields results, which agree fairly well
with experimental heavy ion data in the transverse momentum range from about 
1 GeV to 8 GeV, (see Figs.(4) - (6)).
This agreement supports the basic assumptions of our model:
a) the masses of the
partons have a finite width distribution (Fig.1); b) the momentum distribution 
of partons is described by one and the same function for a wide range of 
transverse momenta;
c) the hadronization happens via effective  quark coalescence;
d) in the hadronization process both  momentum and energy
are nearly conserved locally.  

Finally we note that a spherical
inhomogeneity in the momentum distribution of the quarks leads 
to an azimuthal correlation of hadrons, if they are
produced by the coalescence mechanism described in the
present paper. This effect may be worth of a closer look in future studies.

\section*{Acknowledgments}
This work was supported in part by  Hungarian grants OTKA T34269 and 
T49466. T.S.B. thanks the generous fund
of a Mercator Professorship from the Deutsche Forschungsgemeinschaft
for his sabbatical at the University Giessen.

\end{document}